\documentclass[aps,prl,twocolumn,superscriptaddress,amsmath,amssymb]{revtex4}
\usepackage{graphicx}
\usepackage{color}
\begin{document}
\title{Carrier envelope phase effects in ultrafast, strong-field ionization dynamics of multielectron systems: Xe and CS$_2$}
\author{D. Mathur}
\email{atmol1@tifr.res.in}
\author{K. Dota}
\author{A. K. Dharmadhikari} 
\author{J. A. Dharmadhikari}
\affiliation{Tata Institute of Fundamental Research, 1 Homi Bhabha Road, Mumbai 400 005, India}

\begin{abstract}
Carrier envelope phase (CEP) stabilized 5 fs and 22 fs pulses of intense 800 nm light are used to probe the strong-field ionization dynamics of multielectron entities, xenon and carbon disulfide. We compare ion yields obtained with and without CEP-stabilization: with 8-cycle (22 fs) pulses, Xe$^{6+}$ yields are suppressed (relative to Xe$^+$ yields) by between 30\% and 50\%, depending on phase, reflecting the phase dependence of non-sequential ionization and its contribution to the formation of higher charge states. On the other hand, ion yields for Xe$^{q+}$ ($q$=2-4) with CEP-stablized pulses are {\em enhanced} (by up to 50\%) compared to those with CEP-unstabilized pulses. Such enhancment is particulary pronounced with 2-cycle (5 fs) pulses and is distinctly phase-dependent. Orbital shape and symmetry are found to have a bearing on the response of CS$_2$ to variations in optical field that are effected as CE phase is controllably altered, keeping the overall intensity constant. Molecular fragmentation is found to depend on field strength (not intensity); the observed relative enhancement of fragmentation when CEP-stabilized 2-cycle pulses are used is found to be at the expense of molecular ionization.     
\end{abstract}
\maketitle

Experiments with ultrashort pulses of intense laser light interacting with isolated atoms and molecules continue to invigorate strong-field science (for a recent compilation of cogent reviews, see \cite{PUILS}, and references therein). In such studies, typical intensities of the laser pulses give rise to optical fields whose magnitudes match the intra-molecular Coulombic field. Consequently, the overall interaction is dominated by multiple ionization and, in the case of molecules, inevitably results in the breaking of one or several bonds. In the course of the last decade or so, considerable work carried out using intense pulses of a few tens of femtosecond duration has established the main drivers of the laser-molecule dynamics to be enhanced ionization (EI), spatial alignment, and rescattering ionization \cite{PUILS}. However, the recent availability of few-cycle pulses \cite{wu} has offered tantalizing indications that the dynamics become significantly different when sub-10 fs pulses are used. Dynamic alignment of molecules like O$_2$, N$_2$ no longer occurs as the molecules experience the strong optical field for a period that is far too short for the polarization-induced torque to act on the molecular axis \cite{tong}. Another consequence is that the few-cycle dynamics proceed essentially at equilibrium bond lengths and, consequently, EI is effectively ``switched off" as nuclei do not have sufficient time to move to the critical distance \cite{tong} at which the ionization propensity becomes greatly enhanced. Indeed, Coulomb explosion studies of N$_2$ with 10 fs pulses have confirmed that there is no significant stretching of the N-N bond \cite{baldit}. It may be thought that the dynamics in the few-cycle regime are, therefore, considerably simplified, as only rescattering occurs wherein the electron produced by optical field ionization oscillates in the optical field (on attosecond timescales) and collides with the parent ion, inducing further ionization. Experiments on H$_2$ established that the dynamics are amenable to some measure of control by tuning the intensity and duration of the ultrashort optical field \cite{alnaser2004}. Few-cycle pulses, therefore, offer the prospect of disentangling the different processes that contribute to strong-field molecular dynamics. However, ultrashort dynamics possess even more richness because, as is now being increasingly appreciated, such dynamics are also governed by the instantaneous magnitude of the optical field and not just by the intensity envelope of the incident optical pulse. The parameter of importance, therefore, becomes the carrier envelope phase (CEP), which is a measure of the temporal offset between the maximum of the optical cycle and the maximum of the pulse envelope. The very recent availability of few-cycle pulses whose carrier envelope phase (CEP) can be selected and stabilized opens entirely new vistas for strong field dynamics. A new class of measurements can now be made in which the pulse intensity is kept fixed but the magnitude of the optical field experienced by the irradiated atom or molecule is controlled via the CEP. Reported here are results of experiments that probe the effect that CEP has on the ultrafast dynamics of CS$_2$ and Xe; our results will facilitate the development of new insights into strong field atomic and molecular dynamics in the ultrafast regime and bring to the fore the role of non-sequential ionization in the overall dynamics in the few-cycle regime; we also show that orbital shape and symmetry have a bearing on a molecule's response to variations in optical field that are effected as CE phase is controllably altered, keeping the overall intensity constant. Our experiments with CEP-controlled 5 fs pulses show, counterintuitively, that atomic fragmentation is enhanced and that it depends on the instantaneous strength of the optical field; moreover, the enhancement of fragmentation in the case of phase-stabilized pulses is at the expense of molecular ionization. 

Our target species are both multielectron entities. Xe has, through generation of high harmonics, widespread utility in attosecond science. By measuring ionization spectra of Xe using 5 fs pulses as well as 22 fs pulses we show how CEP affects the formation of Xe$^{q+}$, $q$=2-6. The linear triatomic, CS$_2$ (and its ions), is known to be an important intermediary in chemical transformation processes in cold interstellar plasmas, cometary environments and in planetary and interstellar atmospheres \cite{vardya,cosmovici}. CS$_2$ is also an efficient ionizing agent in charge-exchange organic mass spectrometry \cite{mercer} and has interesting (and important) quantal characteristics in that the four most loosely-bound electrons are in the highest occupied molecular orbital (HOMO) that has pronounced antibonding character (see \cite{CS2paper}, and references therein) which dominates single and multiple ionization. Removal of one, two, or three electrons from the molecule leads to an effective enhancement of the electronic charge density in the internuclear region of CS$_2$, resulting in molecular dications and trications being long-lived (their lifetimes have been measured to be of the order of seconds \cite{aarhus}). Recent work with 4-cycle pulses \cite{CS2paper} showed that molecular ionization dominates the ionization spectrum, with little or no evidence of fragmentation. 
 
Our measurements utilized output pulses from a Ti:Sapphire oscillator which were (i) amplified in a 4-pass amplifier operating at 75 MHz repetition rate, (ii) stretched to $\sim$200 ps and (iii) passed through a programmable acousto-optic dispersive filter that permitted control of pulse shape and duration.  The output passed via an electro-optical modulator (which down-converted to 1 kHz repetition rate) to a 5-pass amplifier and compressor. The resulting 22 fs pulse was further compressed to 5 fs using a 1 m-long Ne-filled hollow fiber and chirped dielectric mirrors. CEP stabilization was accomplished via a fast-loop in the oscillator and a slow-loop in the amplifier \cite{krause}. The extent of CEP stabilization (phase jitter) achieved in the present measurements is depicted in Fig. 1 which also shows a typical interferometric autocorrelation trace. The jitter (typically $<$60 mrad for 22 fs pulses and $<$110 mrad for 5 fs pulses over the course of each measurement) was determined by an $f-2f$ interferometer at a spectrometer acquisition rate of 1 kHz with 920 $\mu$s integration time and 84 ms loop cycle. Laser energy stability with and without CEP stabilization was 0.4\% rms and 1.7\% rms, respectively. Linearly-polarized pulses were transmitted to our molecular beam apparatus through a 300 $\mu$m fused silica window \cite{earlier}; a 5 cm curved mirror was used for focusing down to spot sizes of 7 $\mu$m (width at 1/e$^2$) \cite{supplementary}. Ionization was monitored (with unit collection efficiency) using a linear time-of-flight spectrometer; data acquisition at 1 kHz was in list mode using a 2.5 GHz oscilloscope operating in segmented mode. Figure 1 also depicts the time-evolution of the optical field within each 5 fs pulse for different values of CE phase. Note that although a $\pi$-change in phase results in no change in the waveform other than a reversal of the field's direction, the phase flip manifests itself in the ionization spectra that we present in the following. 

\begin{figure}
\includegraphics[width=6cm]{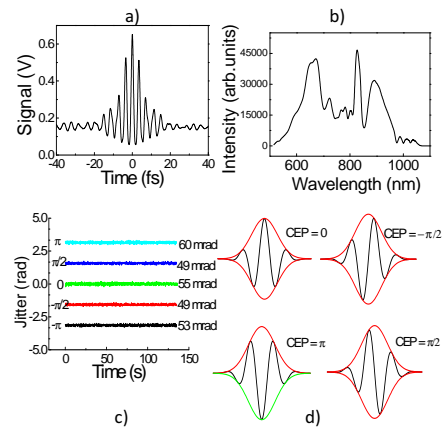}
\caption{Characterization of our 5 fs (2-cycle) laser pulse. a) Time evolution of the pulse and b) its spectral profile; c) Jitter in the CEP stabilization (22 fs) on the timescale of measurements; for 5 fs pulses there is an additional jitter of 60 mrad; and d) Time evolution of the optical field within a single pulse at different CEP values.}
\end{figure}

We measured TOF spectra of Xe-ions at an intensity of 1$\times$10$^{16}$ W cm$^{-2}$ (contrast ratio $>$10$^5$). Our measurements, therefore, were made well in the tunneling regime where the dynamics are entirely optical-field-driven. We note that earlier work on above threshold ionization in Xe with CEP stabilized pulses \cite{Leone} was conducted at lower intensities ($\sim$10$^{13}$ W cm$^{-2}$) where the dynamics are due to a mixture of field-dependent processes and multiphoton ionization (which is dependent upon the intensity of the laser pulse envelope). Typical results obtained for 8-cycle and 2-cycle pulses  are shown in Fig. 2. The striking difference is the observation of charge states up to 6+ in the case of 22 fs pulses and only up to 4+ when 5 fs pulses are used. As discussed below, this is a reflection of non-sequential ionization being suppressed when the number of optical cycles becomes very small. In the following we focus on a comparison of the yields measured for the ratio Xe$^{q+}$/Xe$^{+}$ ($q$=2-6) for CEP-stabilized and CEP-unstabilized pulses of the same peak intensity. That the Xe-ionization spectrum is dominated by field effects is validated in Fig. 2 where we compare how the ratio of Xe$^{q+}$ to Xe$^{+}$ changes as a function of instantaneous field (as expressed in terms of CE phase) for a fixed peak laser intensity for a 2-cycle pulse. Multiple ionization is significantly enhanced (with respect to Xe$^+$ yield obtained with CEP-unstabilized pulses). Figure 2 shows that while there is only marginal enhancement of ion yield for charge states up to 4+, the changes in yield of Xe$^{q+}$, $q$=5,6 are negative and substantial: the yield reduces relative to what is obtained with CEP-unstabilized pulses, possibly because a reduction in the contribution made by non-sequential ionization when CEP-stabilized pulses are used. 

\begin{figure}
\includegraphics[width=6cm]{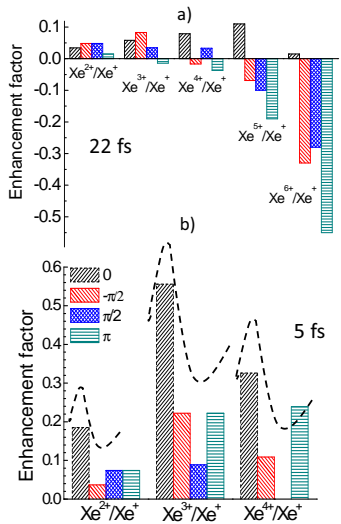}
\caption{Enhancement (with respect to CEP-unstabilized pulses) in the yields of different Xe-ions obtained with different values of CEP-stabilized phase at a peak intensity of 1$\times$10$^{16}$ W cm$^{-2}$ and pulse duration of a) 22 fs and b) 5 fs. The dashed lines depicting CEP-dependent modulation in ion yields are a guide to the eye.}
\end{figure}

It is established that ionization rates for formation of higher charge states of Xe deduced from the oft-used ADK (Ammosov-Delone-Krainov) theory \cite{ADK} are very much less than experimentally measured ones. In the present context, this discrepancy is attributed to the yield of Xe$^{q+}$ ($q$=4-6) ions being mostly due to non-sequential (NS) ionization wherein there is simultaneous tunneling of more than one electron through xenon's field-distorted radial potential function. Yamakawa {\it et al.} \cite{Yamakawa} have shown that in the few-cycle regime there is a suppression of NS ionization. Our results indicate that even with 8-cycle pulses, Xe$^{6+}$ yields are suppressed (relative to Xe$^+$ yields) by between 30\% and 50\%, depending on phase. This reflects the phase dependence of NS ionization and its contribution to the formation of higher charge states. On the other hand, ion yields for Xe$^{q+}$ ($q$=2-4) that we measure with CEP-stablized pulses are actually {\em enhanced} compared to corresponding yields obtained with CEP-unstabilized pulses (Fig. 2). The enhancment is particulary pronounced in the case of 2-cycle pulses and is distinctly phase-dependent, with the largest enhancement being obtained for CEP=0. The relative yields of individual ions exhibits a modulation that depends on the CE-phase. These observation clearly highlight the field-dependent (not intensity-dependent) nature of NS ionization in multielectron atoms like Xe. Rescattering is, of course, one of the drivers of NS ionization \cite{Bhardwaj} and it is, therefore, quite in order that the NS-induced enhancements we observe (Fig. 2) should exhibit a pronounced CEP-dependence (noting that it is the CE-phase that determines when in the course of the optical pulse the ionized electron is ``born"). 
      
In the case of CS$_2$ we make a comparison of the yields measured for the ratio of atomic and molecular ions (with respect to the yield of CS$_2^+$) for CEP-stabilized and CEP-unstabilized pulses of the same peak intensity. As in the atomic case, we find that CEP-stabilized pulses yield an enhancement in the yield of fragment ions and that the relative ion yields exhibits a modulation that depends on the CE-phase. At CEP=0, the atomic fragment signal is very significantly enhanced while the molecular dication and trication yields (with respect to yields obtained with CEP-unstabilized pulses) remain essentially unchanged. At CEP=-$\pi$/2, the relative yield of atomic fragments become more prominent while that of molecular species is reduced (in fact, CS$_2^{3+}$ is no longer seen at this value of CE-phase). The systematic measurements that we have carried out lead us to conclude that atomic fragmentation certainly appears to depend on the instantaneous strength of the optical field. This is seen from the data in Fig. 3 where it is clear that there are differences in relative yields observed for CE phase values of $\pi$/2 and -$\pi$/2. These differences reflect different time evolutions of the optical field for these two phases. Moreover, and significantly, our data offer strong indications that the enhancement of atomic fragments appears to be at the expense of molecular ionization. 

\begin{figure}
\includegraphics[width=6cm]{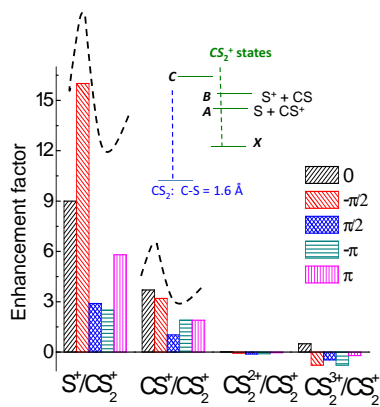}
\caption{Enhancement (with respect to CEP-unstabilized pulses) in the yields of ions obtained from CS$_2$ obtained with different values of CEP-stabilized phase at a peak intensity of 1$\times$10$^{16}$ W cm$^{-2}$ and pulse duration of 5 fs. Note the enhancement of fragment ions S$^+$ and CS$^+$ at the expense of molecular ions CS$_2^{2+}$ and CS$_2^{3+}$. The dashed lines depicting CEP-dependent modulation in ion yields are a guide to the eye. Some electronic states of CS$_2^{q+}$ ($q$=0,1) are also depicted (see text).}
\end{figure}

We note that formation of fragment ions S$^+$ and CS$^+$ by direct ionization of CS$_2$ is not likely as Franck-Condon factors preclude vertical access to the dissociation continua of the ground ($X$) and excited ($A$, $B$) electronic states of CS$_2^+$. The next ionic state, $C$, lies above the dissociation limits S$^+$ + CS and S + CS$^+$ and, hence, fully predissociates. Figure 3 also schematically depicts some electronic states of CS$_2^{q+}$ ($q$=0,1). In long-pulse experiments on CS$_2$, prominent yields of S$^+$ and CS$^+$ fragment ions was accounted for by invoking the enhanced ionization (EI) mechanism wherein the C-S bond length increases so as to allow population of excited electronic states of CS$_2^+$ which are then the precursors of these fragments. The disappearance of these fragments from four-cycle spectra we reported earlier \cite{CS2paper} provided an unambiguous signature that the EI process is ``switched off" in the ultrashort domain. Our observation that the yield of S$^+$ and CS$^+$ fragments with CEP-stabilized pulses is enhanced (compared to the corresponding yield obtained with CEP-unstabilized pulses) indicates that excited electronic states that lie beyond the $C$ state and whose dissociation limits permit formation of S$^+$ and CS$^+$ fragments are being accessed. The electronic configuration in the ground electronic state of CS$_2$ is ${\rm (Core)}^{22}~(5\sigma_g)^2(4\sigma_u)^2(6\sigma_g)^2(5\sigma_u)^2(2\pi_u)^4(2\pi_g)^4$, yielding overall symmetry, $^1\Sigma_g^{~+}$. Ejection of an electron from the $2\pi_u$, $5\sigma_u$, and $6\sigma_g$ orbitals give rise to excited states $A$, $B$, and $C$, respectively. As far as higher excited states are concerned, there is evidence from photoelectron and photionization spectroscopy \cite{MathurHarris} to indicate that these necessitate recourse to consideration of multielectron effects; they are difficult to describe within a single-electron picture but, of course, they are importnt as they manifest the total breakdown of the Koopman's model of ionization that is brought about by very strong final-state correlation effects \cite{Nakatsuji}.
 
In invoking very highly excited electronic states of CS$_2^+$ to validate the enhanced formation of S$^+$ and CS$^+$ fragments that we observe with CEP-stabilized pulses, we note that rescattering is unlikely to contribute in this instance. In earlier work with 11 fs CEP-unstabilized pulses it has been established that doubly- and triply-charged molecular ions (CS$_2^{2+}$, CS$_2^{3+}$) dominate the four-cycle spectrum, ostensibly at the expense of fragmentation channels \cite{CS2paper}. This was taken to be a signature of rescattering being ``switched off" because of constraints imposed by the quantum-mechanical nature of CS$_2$'s outermost antibonding $2\pi_g$ orbital. The wavepacket of the returning electron interferes destructively with the spatial extent of this orbital, leading to effective cancellation of the rescattering process. The returning electron's energy is, consequently, no longer available for electronic excitation to high-lying CS$_2^{+*}$ states that are quantally allowed to dissociate into S$^+$+CS$^+$. As far as the phase effects seen in Fig. 3 are concerned, they manifest how the individual molecular orbitals, $2\pi_g$, $2\pi_u$, $5\sigma_u$, and $6\sigma_g$ respond to the field variations that are experienced with the 5 fs pulse under different CEP conditions. Nonperturbative time-dependent density functional theory (TD-DFT) has been applied to numerically solve Kohn-Sham equations for CS$_2$ exposed to laser intensities in excess of 10$^{14}$ W cm$^{-2}$ \cite{Bandrauk} and the results do, indeed, verify that different field magnitudes affect each orbital (including inner orbitals) in different fashion. We note that the symmetry of individual orbitals will play a role in rationalizing the observations of Fig. 3 \cite{faisal,n2o2} on how relative ion yields are affected by different values of CEP phase. The $\pi$-orbitals have a nodal plane containing the molecular axis and, consequently, will make a lower contribution to the ionization yield when aligned parallel to the laser's polarization vector. On the other hand, orbitals with $\sigma$ symmetry will ionize most effectively as their density is maximum parallel to the optical field.      

Experiments on CS$_2$ that we conducted with 22 fs CEP-unstabilized pulses yielded data that reproduce well earlier results obtained with CEP-unstabilized four-cycle pulses \cite{CS2paper}, with molecular ionization overwhelmingly dominating the dynamics and a drastic reduction in contribution from atomic fragments (compared to measurements made with pulses of 50 fs and longer duration). It is, therefore, the CEP phase in our 5 fs pulses that drives the dynamics depicted in Fig. 3.
            
Intense few-cycle pulses within which the optical field can be precisely fixed via CEP control will open new opportunities for controlling both the moment when an electron wavepacket is ``born" and its subsequent motion. This capability provides a new flip to attosecond science. Enhancement of the intensity of such CEP-stabilized pulses will permit control of electronic motions in the inner-orbitals, enabling new classes of experiments to be conducted on heavy atoms (like Xe) and molecules containing heavy atoms (like CS$_2$) in which such electrons are relativistic. Furthermore, little is known about possible interplay of electrons in inner and outer orbitals in multielectron entities. There has been inconclusive debate on how effectively external fields may be shielded from electrons in inner orbitals \cite{L'Huillier}. Screening effects in entities like Xe and CS$_2$ would make it difficult to estimate the local field experienced by inner valence electrons, making systematic descriptions of the dynamics in multielectron systems an intractable theoretical problem. The observations we have presented here should aid in testing the efficacies of future theoretical developments in this direction. From the perspective of molecules, those that comprise heavy atoms (like CS$_2$) require dipole and polarizability corrections to be incorporated into existing tunneling and other strong-field theories; such corrections need to be CEP-dependent and, as in the case of heavy atoms, it is anticipated that the present results will stimulate further theoretical work.


\begin{thebibliography}{00}

\bibitem{PUILS}K. Yamanouchi, {\it et al.}, Progress in Ultrafast Intense Laser Science, Vols. 1-9 (Berlin: Springer) 2010.
\bibitem{wu}Z. Wu, {\it et al.}, J. Chem. Phys. {\bf 126}, 074311 (2007). 
\bibitem{tong}X. M. Tong, {\it et al.}, J. Phys. B {\bf 38}, 333 (2005).
\bibitem{baldit}E. Baldit, S. Saugout, and C. Cornaggia, Phys. Rev. A {\bf 71}, 021403 (2005). 
\bibitem{alnaser2004}A. S. Alnaser, {\it et al.}, Phys. Rev. Lett. {\bf 93}, 183202 (2004), {\it ibid.} {\bf 93}, 113003 (2004).
\bibitem{vardya}M.S. Vardya and S.P. Tarafdar in Astrochemistry (D. Reidel, Dordrecht), p.604 (1987).
\bibitem{cosmovici}Cosmovici {\it et al}, Nature, {\bf 310}, 122 (1984), Astron. Astrophys., {\bf 114}, 373 (1982). 
\bibitem{mercer}Mercer {\it et al}, Org. Mass. Spectrom., {\bf 21} 717 (1986), Dass {\it et al}, {\it ibid.}, {\bf 21}, 741 (1986).
\bibitem{CS2paper}D. Mathur, A. K. Dharmadhikari, F. A. Rajgara, and J. A. Dharmadhikari, Phys. Rev. A. {\bf 78}, 013405 (2008).
\bibitem{aarhus}D. Mathur, {\it et al}, J. Phys. B {\bf 28}, 3415 (1995).
\bibitem{krause}J. Rauschenberger {\it et al.}, Laser Phys. Lett. {\bf 3}, 37 (2006).
\bibitem{earlier}D. Mathur, A. K. Dharmadhikari, F. A. Rajgara, and J. A. Dharmadhikari, Phys. Rev. A {\bf 78}, 023414 (2008), A. K. Dharmadhikari, J. A. Dharmadhikari, F. A. Rajgara, and D. Mathur, Opt. Express {\bf 16}, 7083 (2008), F. A. Rajgara, D. Mathur, A. K. Dharmadhikari, and C. P. Safvan, J. Chem. Phys. {\bf 130}, 231104 (2009).
\bibitem{supplementary}See Fig. S1 in the Supplemental Material. The focused beam was magnified with a 40x microscope objective and imaged on to a 12-bit CCD camera (pixel size 4.4 $\mu$m). 
\bibitem{Leone}M. J. Abel, {\it et al}, J. Phys. B {\bf 42}, 075601 (2009).
\bibitem{ADK}M. V. Ammosov {\it et al.}, Zh. Eksp. Teor. Fiz. {\bf 91}, 2008 (1986) [Sov. Phys. JETP {\bf 64}, 1191 (1986)].
\bibitem{Yamakawa}K. Yamakawa, {\it et al}, Phys. Rev. Lett. {\bf 92}, 123001 (2004).
\bibitem{Bhardwaj}V. R. Bhardwaj, {\it et al.}, Phys. Rev. Lett. {\bf 86}, 3522 (2001).
\bibitem{MathurHarris}D. Mathur and F. M. Harris, Mass Spectrom. Rev. {\bf 8}, 269 (1989), and references therein.
\bibitem{Nakatsuji}H. Nakatsuji, Chem. Phys. {\bf 76}, 283 (1983).
\bibitem{Bandrauk}E. P. Fowe and A. D. Bandrauk, Ch. 21 in Coherence and Ultrashort Pulse Laser Emission, Ed. F. J. Duarto (Vienna: Intech) 2011.
\bibitem{faisal}J. Muth-B\"ohm, A. Becker, and F. H. M. Faisal, Phys. Rev. Lett. {\bf 85}, 2280 (2000).
\bibitem{n2o2}M. Okunishi, K. Shimada, G. Pr\"umper, D. Mathur, and K. Ueda, J. Chem. Phys. {\bf 127}, 064310 (2007).
\bibitem{L'Huillier}A, L'Huillier, L. Jonsson, and G. Wendin, Phys. Rev. A {\bf 33}, 3938 (1986), A. Szoke and C. K. Rhodes, Phys. Rev. Lett. {\bf 56}, 720 (1986), and references therein.











\end{thebibliography}
\end{document}